\documentclass[12pt,a4paper]{article}\usepackage{amssymb}\usepackage{bezier}\usepackage{emlines2}\usepackage[T2A]{fontenc}\usepackage[cp866]{inputenc}\usepackage[russian,english]{babel}
\normalbaselines
\begin{document}
\voffset=-2cm
\begin{center}
{\LARGE {\bf {Black holes: interfacing the classical \\[3mm]
and the quantum}}}\footnote{The talk at the
International School/Seminar {\it Quantum Field Theory and Gravity},
July 2--7, 2007, Tomsk, Russia} \\[5mm]
{\bf B.\ P.\ Kosyakov} \\[3mm]
{Russian Federal Nuclear Center, Sarov, 607190 Nizhnii Novgorod Region, 
Russia} \\[3mm] {\it E-mail}: \verb+kosyakov@vniief.ru+
\end{center}

\begin{abstract}
\noindent
The central idea advocated in this paper is
that
{forming the black hole horizon  is attended with 
transition from the classical regime of evolution to the quantum one}. 
We justify the following
criterion for discriminating
between the
classical and the quantum:
{spontaneous creations and annihilations of 
particle-antiparticle pairs 
are impossible in the classical world but possible in the quantum
world}.
We show that
it is sufficient to 
{change the overall sign of the
spacetime signature in the classical picture of field propagation for
it to be treated as its associated quantum picture}.
To describe a self-gravitating object
at 
the last stage of its classical evolution, 
we propose to use the
Foldy--Wouthuysen 
representation of the Dirac equation in curved spacetimes,
and the Gozzi classical path integral.
In both approaches, maintaining the dynamics in the classical regime 
is controlled by supersymmetry.
\end{abstract}

\section{Introduction}
The existence of black holes is one of the most intriguing predictions of
general relativity (for a review see, e.g., \cite{Wheeler, Chandrasekhar}). 
A black hole is defined as a region of spacetime from behind which it is
impossible to 
escape 
to the future null infinity ${\cal I}^+$ 
without exceeding the speed of light.
This region is invisible to distant observers.
There is a boundary, called the event horizon, between an exterior region, where it 
is possible for signals to reach infinity, and 
the interior region, where signals remain trapped. 
Astrophysical black holes are assumed to arise as the final states of 
stellar evolution of sufficiently massive stars.

The formation of a black hole resembles a {\it phase transition} in
condensed matter physics.
The approach to black-hole states occurs fairly {quickly}.
For a Schwarzschild black hole, 
the time scale for this approach is estimated at
\begin{equation}
\tau=\frac{c}{\kappa}
=\frac{2r_{\rm S}}{c}\,,
\label
{time-scale-BH-formation}
\end{equation}
where $\kappa$ is the surface gravity $\kappa=c^4/(4GM)$, 
and $r_{\rm S}=2GM/c^2$ is
the Schwarzschild radius.
Numerically, $\tau\simeq 2\cdot 10^{-5}(M/M_\odot)$ s, where $M_\odot$ is the mass 
of the Sun.
If  $M\sim M_\odot$, then $\tau\sim  10^{-5}$ s.

The {\it energy content} of the system in the initial state differs 
significantly from that in the final
state.
To see this, we refer to the Penrose process for extracting energy from a
rotaing and/or charged black hole.
For the  Kerr--Newman metric,  
\begin{equation}
M^2
=
\left(M_{\rm irr}
+\frac{Q^2}{4GM_{\rm irr}}\right)^2
+
\frac{J^2}{4G^2M^2_{\rm irr}}\,,
\label
{M-irr}
\end{equation}
where $M$ is the total mass of the black hole, and $M_{\rm irr}$ is its
irreversible mass.
The maximum amount of rotational energy which can be 
extracted from an extreme Kerr black hole prior to slowing down its rotation
is approximately $29 {\%}$ of its total energy. 
Furthermore, about  $50 {\%}$ can be stored in the form of electromagnetic
energy.
By contrast,
only few percent of nuclear binding energy can be 
emitted by stars 
throughout their lifetimes.

This transition entails a profound {\it symmetry 
rearrangement}.
By the `no hair theorems' \cite{Heusler, Chrusciel}, 
each isolated, stationary black hole
is completely described  by three parameters: its mass $M$, 
angular momentum $J$, and 
electric charge $Q$.
Whatever the structure of a star which collapses under its own 
gravitational field, 
the exterior of
the resulting black hole is described by a Kerr--Newman 
solution.
In other words, all initial symmetries of the collapsing system 
and their associated conservation
laws, except for $M$, $J$, and $Q$, disappear in
the ultimate black-hole state. 

An appreciable distinction between 
this phenomenon and ordinary
phase transitions  
lies in its {\it irreversibility}. 
Once converted into a black hole, the system
can never regain its previous  state. 

What is the nature of this phase transition?
If we take as our basic paradigm a collapsing star which settles down to
a stationary black hole,
then a plausible assumption is that 
{\bf forming the black hole horizon is a
transition between the classical and quantum regimes of evolution}. 
Indeed, the initial incarnation of this system is definitely classical.
On the other hand, 
the resulting black hole is a quantum object.
Any black hole evaporates due to Hawking radiation.
\cite{Hawking4,Hawking5}. 
We now refer to
the spontaneous creation of particle-antiparticle pairs in a strong
gravitational field near the black hole horizon.
One member of a 
virtual pair 
can fall into the black hole while its partner
escapes to infinity.
We recognize Hawking radiation in these processes of pair creation and the
subsequent particle escaping. 

According to the Hawking--Penrose singularity theorems \cite{HawkingEllis}, 
the 
collapse terminates in a singularity 
(big crunch) where the curvature is
infinite and the classical concepts of space and time lose their meaning.  
However, quantum-gravitational effects are expected to be 
dominant as one approaches the Planck scale $\ell_{\rm P}=\sqrt{G\hbar/c^3}\simeq 1.6\cdot
10^{-33}$ cm.
Taking into account the impact of trans-Planckian modes 
radiated outward from the center, one can infer that 
the 
interior 
of the black hole is governed by the laws of quantum physics.

Therefore, a collapsing system runs through two phases.
The initial phase, before
the black hole has formed, may be defined as classical, while the final 
phase, after the hole has formed, may be defined as quantum.

This paper is organized as follows.
In Sec.~2 we obtain two criteria for discriminating
between the
classical and the quantum.
We then argue that the black hole horizon is just the geometrical layout required
to interface the classical and quantum realms. 
To take a closer look at this demarcation, we should have a general 
strategy together with suitable techniques.
In Sec.~3 we briefly review the Foldy--Wouthuysen approach to
the description of a Dirac particle, which provides 
a convenient representation of the classical regime of evolution.
We will see that the Foldy--Wouthuysen transformation is possible in
some curved spacetimes with stationary geometries, specifically in 
the Schwarzschild manifold.
Furthermore, the feasibility of an exact 
Foldy--Wouthuysen transformation for a given system turns out to be related to 
the existence of the supersymmetry properties of this system.
In Sec.~4 we outline the classical path integral concept. 
We demonstrate that the classical path integral can be obtained from the 
Feynman path integral if we change time $t$ for the
`supertime' $(t,\theta,{\bar\theta})$, and the phase space 
coordinates $q$ and $p$ for
the super-phase space coordinates  $Q$ and $P$.
This structure of the classical path integral makes it clear that
the classical--quantum phase transition is attended with supersymmety 
violation.
Section 5 summarizes our discussion.

\section{The classical and the quantum}
What is the difference between the classical and quantum 
views of the same dynamical entity?
We address this question by comparing the properties of {\it particles} 
and {\it fields}
in the classical and quantum descriptions.
A convenient framework bringing together the classical and 
quantum  treatments is provided by the {path integral} approach.

Let us begin with particles.
A quantum-mechanical particle can be described by the Feynman path integral
\begin{equation}
K({\bf z}_f, T|{\bf z}_i,0)=\int [{{\cal D}}q]\,\exp\left[\frac{i}{\hbar}\int_0^T dt\,
L(q,q')\right].
\label
{Feynman-PI}
\end{equation}                                           
Here, $[{{\cal D}}q]$ means integration is to be carried out in the 
space of all paths from ${\bf z}_i(0)$ to ${\bf z}_f(T)$.

\begin{figure}[htb]
\begin{center}
\unitlength=1.00mm
\special{em:linewidth 0.4pt}
\linethickness{0.4pt}
\begin{picture}(53.00,60.00)
\put(0.00,15.00){\vector(0,1){28.00}}
\put(0.00,48.00){\makebox(0,0)[cc]{\tt time}}
\put(53.00,55.00){\makebox(0,0)[cc]{$x^\mu=(T,{\bf z}_f)$}}
\put(30.00,5.00){\makebox(0,0)[cc]{$x^\mu=(0,{\bf z}_i)$}}
\put(25.00,10.00){\makebox(0,0)[cc]{$\bullet$}}
\put(45.00,50.00){\makebox(0,0)[cc]{$\bullet$}}
\bezier{192}(25.00,10.00)(29.00,38.00)(45.00,50.00)
\bezier{120}(25.00,10.00)(14.00,21.00)(15.00,35.00)
\bezier{220}(15.00,35.00)(17.00,60.00)(45.00,50.00)
\bezier{104}(45.00,50.00)(37.00,37.00)(45.00,30.00)
\bezier{164}(45.00,30.00)(63.00,6.00)(45.00,15.00)
\bezier{64}(45.00,15.00)(36.00,21.00)(32.00,18.00)
\bezier{48}(32.00,18.00)(26.00,12.00)(25.00,10.00)
\end{picture}
\caption{World lines contributing to the Feynman path integral}
\label
{paths}
\end{center}
\end{figure}
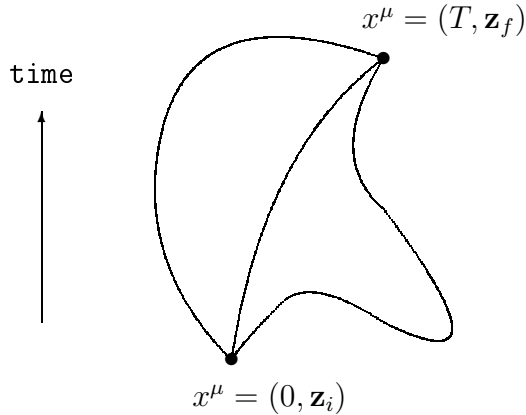

Whatever the kind of the world line $z^\mu(\tau)$ 
passing through the
points  $x^\mu=(0,{\bf z}_i)$ and  $x^\mu=(T,{\bf z}_f)$
(as in Figure \ref{paths}), it makes a contribution to the Feynman path 
integral
provided that the Lagrangian $L$ is real, and expression
(\ref{Feynman-PI}) is well defined for
this path.

To illustrate,  the Poincar\'e--Planck  Lagrangian
\begin{equation}
L=-m\,\sqrt{{\dot z}^2} 
\label
{P-P-Lagrangian}
\end{equation}                                           
is real and finite only for timelike paths.
If $z^\mu(\tau)$ is a null curve, then $L=0$.
If $z^\mu(\tau)$ is spacelike, then $L$ is complex-valued. 
Since the imaginary part of $L$ can take both positive and negative values,
expression (\ref{Feynman-PI}) is ill-defined.  
By contrast, the Lagrangian proposed by Brink--Deser--Zumino--Di Vecchia--Howe 
\cite{Brink}
\begin{equation}
{L}=-\frac{1}{2}\left(\eta\,{\dot z}^{2}+\frac{ m^{2}}{\eta}\right)
\label
{Brink-lagrangian}
\end{equation}
is real and finite
for timelike, null, and spacelike curves.
Here $\eta$ is an
auxiliary dynamical variable, sometimes called {einbein}.

We now look at $\Lambda$- and $V$-shaped world lines.
Let a particle be moving along a timelike world line 
from the remote past to the future up to the point ${\rm A}$, 
and then returns to the remote past.
This $\Lambda$-shaped world line of a {single particle} can be interpreted as
that representing the annihilation of a pair  that occurs at a point ${\rm A}$, 
because the antiparticle of this particle 
may be thought of as an 
object identical to it in every respect but moving back in
time.
\begin{figure}[htb]
\begin{center}
\begin{minipage}[t]{78mm}
\unitlength=1.00mm
\special{em:linewidth 0.4pt}
\linethickness{0.4pt}
\begin{picture}(111.00,45.00)
\put(10.00,10.00){\vector(0,1){6.00}}
\put(27.00,16.00){\vector(0,-1){6.00}}
\bezier{40}(10.00,16.00)(10.00,22.00)(12.00,26.00)
\bezier{52}(18.00,38.00)(14.00,30.00)(12.00,26.00)
\put(18.00,38.00){\makebox(0,0)[cc]{$\bullet$}}
\put(18.00,43.00){\makebox(0,0)[cc]{$\rm A$}}
\put(52.00,36.00){\vector(0,1){7.00}}
\put(71.00,43.00){\vector(0,-1){7.00}}
\bezier{56}(71.00,36.00)(71.00,29.00)(69.00,22.00)
\bezier{56}(69.00,22.00)(66.00,13.00)(63.00,10.00)
\bezier{76}(63.00,10.00)(55.00,22.00)(54.00,26.00)
\bezier{40}(54.00,26.00)(52.00,32.00)(52.00,36.00)
\put(63.00,10.00){\makebox(0,0)[cc]{$\bullet$}}
\put(63.00,6.00){\makebox(0,0)[cc]{$\rm B$}}
\bezier{56}(18.00,38.00)(19.00,33.00)(25.00,24.00)
\bezier{32}(25.00,24.00)(27.00,20.00)(27.00,16.00)
\end{picture}
\caption{$\Lambda$- and $V$-shaped paths}
\label{crann}
\end{minipage}
\end{center}
\end{figure}
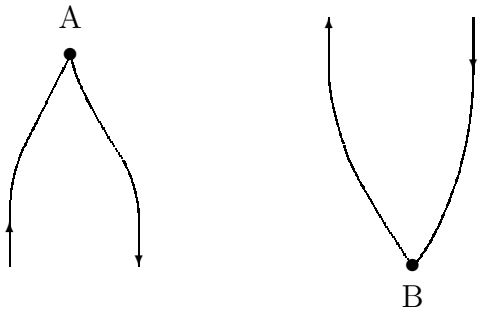
Likewise, given a $V$-shaped world line of a {single particle} 
that runs initially from the far future to the 
past up to the point ${\rm B}$ and then returns to the far future, 
we interpret it as that representing 
the birth of a pair occurring at a point ${\rm B}$.
Any $\Lambda$- or $V$-shaped world line 
passing through the chosen end
points  $(0,{\bf z}_i)$ and  $(T,{\bf z}_f)$ contributes 
to the Feynman path integral (\ref{Feynman-PI}).
The quantum description leaves room for both particles and
antiparticles together with their creations and annihilations.

On the other hand, 
classical particles are governed by the principle of least action.
This principle 
can be formulated
for smooth timelike and null world 
lines.
However, it defies unambiguous formulation for 
$V$- and $\Lambda$-shaped world lines.
Indeed, consider a spacelike hyperplane which represents space at some
instant in a particular Lorentz frame.
This hyperplane intersects a 
$\Lambda$-shaped curve twice, otherwise it fails to intersect this curve 
at all. 
The same is true for 
$V$-shaped curves.
Thus, although the classical picture allows the coexistence of particles 
and antiparticles,  creations and annihilations of pairs, represented by  
$V$- and $\Lambda$-shaped world lines, are banned \cite{k06}.

Therein lies the fundamental difference between the quantum and classical 
viewpoints on
particles: {\bf spontaneous creations and annihilations of pairs are 
permissible in the quantum
reality and impermissible in the classical reality} 
\footnote{It may be worth
pointing out that behind this criterion is the fact that quantum 
theory gets by with a well defined {action} (appearing in the Feynman 
path integral) while classical 
theory requires additionally that the action be suited to 
its extremization following Hamilton's principle.}.

Hawking radiation, associated with the pair creation in a strong 
gravitational field near the black hole horizon, 
is a characteristically quantum phenomenon. 

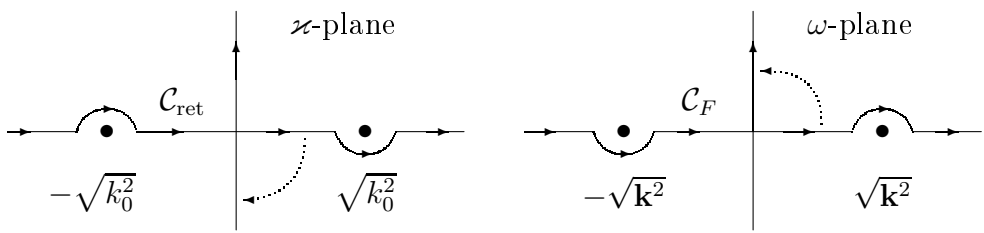
\begin{figure}[htb]
\begin{center}
\unitlength=1.00mm
\special{em:linewidth 0.4pt}
\linethickness{0.4pt}
\begin{picture}(148.00,38.00)
\put(6.00,17.00){\vector(1,0){3.00}}
\emline{9.00}{17.00}{1}{15.00}{17.00}{2}
\put(19.00,20.00){\vector(1,0){1.00}}
\put(23.00,17.00){\vector(1,0){6.00}}
\emline{29.00}{17.00}{5}{40.00}{17.00}{6}
\bezier{20}(15.00,17.00)(16.00,20.00)(19.00,20.00)
\bezier{16}(19.00,20.00)(22.00,20.00)(23.00,17.00)
\put(19.00,17.00){\makebox(0,0)[cc]{$\bullet$}}
\put(36.00,24.00){\vector(0,1){5.00}}
\emline{36.00}{29.00}{5}{36.00}{33.00}{6}
\put(104.00,17.00){\vector(0,1){12.00}}
\emline{104.00}{29.00}{5}{104.00}{33.00}{6}
\put(17.00,9.00){\makebox(0,0)[cc]{$-\sqrt{k_0^2}$}}
\put(53.00,9.00){\makebox(0,0)[cc]{$\sqrt{k_0^2}$}}
\put(87.00,9.00){\makebox(0,0)[cc]{$-\sqrt{{\bf k}^2}$}}
\put(121.00,9.00){\makebox(0,0)[cc]{$\sqrt{{\bf k}^2}$}}
\put(29.00,21.00){\makebox(0,0)[cc]{${\cal C}_{\rm ret}$}}
\put(97.00,21.00){\makebox(0,0)[cc]{${\cal C}_{F}$}}
\put(50.00,31.00){\makebox(0,0)[cc]{$\varkappa$-plane}}
\emline{36.00}{4.00}{7}{36.00}{24.00}{8}
\put(40.00,17.00){\vector(1,0){3.00}}
\put(106.00,25.00){\vector(-1,0){1.00}}
\bezier{16}(106.00,25.00)(113.00,25.00)(113.00,18.00)
\put(38.00,8.00){\vector(-1,0){1.00}}
\bezier{16}(38.00,8.00)(45.00,9.00)(45.00,16.00)
\emline{43.00}{17.00}{9}{49.00}{17.00}{10}
\put(53.00,14.00){\vector(1,0){1.00}}
\emline{57.00}{17.00}{13}{61.00}{17.00}{14}
\put(61.00,17.00){\vector(1,0){3.00}}
\emline{63.00}{17.00}{17}{66.00}{17.00}{18}
\bezier{16}(49.00,17.00)(50.00,14.00)(53.00,14.00)
\bezier{20}(53.00,14.00)(56.00,14.00)(57.00,17.00)
\put(53.00,17.00){\makebox(0,0)[cc]{$\bullet$}}
\put(118.00,31.00){\makebox(0,0)[cc]{$\omega$-plane}}
\put(108.00,17.00){\vector(1,0){4.00}}
\emline{112.00}{17.00}{19}{117.00}{17.00}{20}
\put(121.00,20.00){\vector(1,0){1.00}}
\emline{125.00}{17.00}{21}{128.00}{17.00}{22}
\put(128.00,17.00){\vector(1,0){3.00}}
\emline{131.00}{17.00}{23}{134.00}{17.00}{24}
\emline{108.00}{17.00}{25}{97.00}{17.00}{26}
\put(94.00,17.00){\vector(1,0){3.00}}
\emline{95.00}{17.00}{27}{91.00}{17.00}{28}
\put(87.00,14.00){\vector(1,0){1.00}}
\emline{83.00}{17.00}{29}{78.00}{17.00}{30}
\put(74.00,17.00){\vector(1,0){4.00}}
\emline{76.00}{17.00}{33}{74.00}{17.00}{34}
\bezier{16}(117.00,17.00)(118.00,20.00)(121.00,20.00)
\bezier{20}(121.00,20.00)(124.00,20.00)(125.00,17.00)
\bezier{16}(91.00,17.00)(90.00,14.00)(87.00,14.00)
\bezier{20}(87.00,14.00)(84.00,14.00)(83.00,17.00)
\put(87.00,17.00){\makebox(0,0)[cc]{$\bullet$}}
\put(121.00,17.00){\makebox(0,0)[cc]{$\bullet$}}
\emline{104.00}{4.00}{7}{104.00}{17.00}{8}
\end{picture}
\caption{The integration contours ${\cal C}_{\rm ret}$ and 
${\cal C}_{F}$ suitable for ${\widetilde D}_{\rm ret}$ 
and ${\widetilde D}_F$}
\label
{bounbary}
\end{center}
\end{figure}

We now turn to fields.
Having in mind the path integral approach, 
the key difference 
between the classical and quantum manifestations of the same field is due to 
the different boundary conditions 
imposed on their propagation laws.
To be specific,  consider a massles scalar field. 
The Fourier transform of the 
retarded Green's function
\begin{equation}
{\widetilde D}_{\rm ret}(k)
=-\frac{1}{k^2+2ik_0\epsilon}
=\frac{1}{{\bf k}^2-(k_0+i\epsilon)^2}
\label
{retarded-propagator}
\end{equation}
gives an accurate account of how this field 
propagates in  classical theory.
If the integration over the variable  $\varkappa=|{\bf k}|$ is carried out 
first\footnote{We suppose that angular variables, if
any, have already been integrated out, and that the resulting expression can
be uniquely continued to the negative  $\varkappa$-semiaxis.}, then the poles at
\begin{equation}
\varkappa=\pm\sqrt{k_0^2} \pm i\epsilon
\label
{retarded-propagator-poles}
\end{equation}
are avoided by the path depicted in the left plot of Figure \ref{bounbary}.
Indeed, the pole with the positive real part lies above the integration path
and the pole with the negative real part lies under  the path.
If the poles approach the real axis, the integration path should be
slightly deformed to form the curve ${\cal C}_{\rm ret}$ in the complex 
$\varkappa$-plane.

The propagation of a free massles field in quantum theory is 
described by the Feynman propagator 
\begin{equation}
{\widetilde D}_{F}(k)=-\frac{1}{k^2+i\epsilon}\,,
\label
{Feynman-propagator}
\end{equation}
which obeys the causal boundary condition.
It follows the prescription for avoiding the poles 
\begin{equation}
\omega=\pm\sqrt{{\bf k}^2} \mp i\epsilon,
\label
{Feynman-propagator-poles}
\end{equation}
where $\omega$ denotes $k_0$.
The integration contour ${\cal C}_{F}$ in the complex $\omega$-plane is 
depicted in the right plot of Figure \ref{bounbary}.
Exact propagators of 
interacting quantum fields are given by the
spectral K{\"a}ll\'en--Lehmann representation
obeying the same boundary condition.

In order to bridge the gap between the retarded and causal boundary conditions, 
we  `euclideanize' both descriptions.

Assuming that the integrand decreases sufficiently fast as $\varkappa$ 
approaches
infinity,  
it is possible to rotate the path of integration  ${\cal C}_{\rm ret}$ 
in a clockwise direction by $\frac{\pi}{2}$ in the complex $\varkappa$-plane 
without crossing the poles, as in left plot of Figure \ref{bounbary}. 
This operation, the analytical continuation to imaginary values of the complex 
$\varkappa$-plane,
is similar to the {Wick rotation}. 
Introducing a new variable ${\mathbb K}=i{\bf k}$ makes the length squared of 
$k^\mu$
positive definite: 
\begin{equation}
k^2_E=k_0^2+{\mathbb K}^2. 
\label
{Euclid-class}
\end{equation}
The analytical continuation of the space variables to the imaginary axes 
\begin{equation}
{\mathbb X}=i{\bf x},
\label
{x=-iX}
\end{equation}                                          
performed together with the analytical continuation in 
${\bf k}$-space, introduces the
Euclidean metric
\begin{equation}
dx^2_E=dx_0^2+d{\mathbb X}^2.
\label
{x-sqr-E=x-4-sqr+-bf-x-sqr}
\end{equation}                                          

On the other hand, if we carry out the Wick rotation of 
the path of integration  ${\cal C}_{F}$ 
in a counterclockwise direction by $\frac{\pi}{2}$ in the complex 
$\omega$-plane 
without crossing the poles, as in right plot of Figure \ref{bounbary}, which is
equivalent to introducing ${k}_4=i{k}_0$, then the length squared of 
$k^\mu$ becomes
negative definite: 
\begin{equation}
k^2_E=-\left(k_4^2+{\bf k}^2\right). 
\label
{Euclid-quant}
\end{equation}
The analytical continuation of the time variable to the imaginary axis 
\begin{equation}
{x}_4=-i{x}_0,
\label
{x-0=-ix-4}
\end{equation}                                          
performed together 
with the Wick rotation, introduces the
Euclidean metric
\begin{equation}
dx^2_E=-\left(dx_4^2+d{\bf x}^2\right).
\label
{x-sqr-E=x-0-sqr+bf-x-sqr}
\end{equation}                                          

We thus see that it is sufficient to 
{\bf change the overall sign of the
spacetime signature in the classical description of field propagation for
it to be treated as the quantum description of field propagation}.
Indeed, two Lorentzian metrics with opposite signatures can always
be analytically
continued to two Euclidean line elements of opposite sign, such as those shown in 
(\ref{x-sqr-E=x-4-sqr+-bf-x-sqr}) and 
(\ref{x-sqr-E=x-0-sqr+bf-x-sqr}).

Taken alone, the overall sign of the Lorentzian
metric is of no particular importance,
its choice is 
a matter of convention.
In fact, the HEP theorists prefer the mainly negative signature $(+---)$,
while the
relativists like the mainly positive signature
$(-+++)$.  
However, if this overall sign is changed as one passes from some region of 
spacetime to a
contiguous region, then this change of sign is evidence of switching 
between the classical 
and quantum regimes of field propagation.

Such is the case for the contiguous regions 
inside and outside the 
event horizon of a black hole.
As the simplest example, we refer to the Schwarzschild metric describing an isolated
spherically symmetric stationary black hole
\begin{equation}
ds^2=\left(1-\frac{r_{\rm S}}{r}\right)
dt^2- 
\left(1-\frac{r_{\rm S}}{r}\right)^{-1}
dr^2-r^2d{\Omega}.
\quad
\label
{Schwarzschild-metric}
\end{equation}                                          
Here, $d{\Omega}$ is the round metric in $S^2$, and $r_{\rm S}
={2GM}/{c^2}$ is the Schwarzschild radius which represents the event 
horizon of this black hole. 
In the 
Schwarzschild exterior $r>r_{\rm S}$, 
the Killing vector field $X=\partial_t$ is interpreted as the asymptotic time
translation. 
In the Schwarzschild interior $r<r_{\rm S}$, 
$r$ is a time coordinate, and the integral lines 
of the vector field $X=\partial_r$ are incomplete timelike geodesics which
terminate at $r=0$.
Once the Euclideanization has been performed, the 
regions inside and outside the boundary $r=r_{\rm S}$
take the Euclidean
metrics of the type of 
(\ref{x-sqr-E=x-4-sqr+-bf-x-sqr}) and (\ref{x-sqr-E=x-0-sqr+bf-x-sqr}), 
respectively.
The question now arises of what happens to the physical
reality at the surface of the collapsing star
when the Schwarzschild radius $r=r_{\rm S}$ is crossed.
Does the classical picture give way to the quantum picture, say,
all cats which populate this star become Schr\"odinger's cats? 

It is well known that the surface $r=r_{\rm S}$ is locally
perfectly regular.
The singularity at $r=r_{\rm S}$ is a mere coordinate singularity
in the original  Schwarzschild coordinate frame.
In some other coordinates (such as Kruskal--Szekeres coordinates), the metric 
is
smooth at $r=r_{\rm S}$. 
When crossing $r=r_{\rm S}$, 
an observer on the surface of a collapsing star 
 will feel no geometrical and dynamical jumps.
However, globally, $r=r_{\rm S}$ acts as a pont of no return.
Furthermore, 
every light
cone tilts over at this point, so that 
the roles of $t$ and $r$ are interchanged---no matter what the coordinate frame is used.  
Therefore, the phase transition  
related to forming the Schwarzschild black hole
should be viewed as a global concept: the entire spacetime must be 
known before its existence and form can be determined. 

Consider a rotating black hole.
The Kerr
metric in Boyer--Lindquist coordinates reads
\begin{eqnarray}
ds^2
=
\frac{\Delta-a^2\sin^2\vartheta}{\Sigma}\,dt^2
+ 
\frac{2a\sin^2\vartheta\left(r^2+a^2-\Delta\right)}{\Sigma}\,dt\,d\varphi 
\nonumber\\
-\frac{\left(r^2+a^2\right)^2-\Delta\,a^2\sin^2\vartheta}{\Sigma}\,
\sin^2\vartheta\,
d\varphi^2
-
\frac{\Sigma}{\Delta}\,dr^2
-{\Sigma}\,d\vartheta^2,
\label{Kerr-metric}\end{eqnarray}                                          
where ${\Sigma}=r^2+a^2\cos^2\vartheta$, ${\Delta}=r^2+a^2-2GMr/c^2$.
If $0\le a<GM/c^2$, then there are two distinct event horizons at 
$r=r_\pm$ where $r_\pm$ 
are the roots of $\Delta=0$,
\begin{equation}
\Delta
=r^2+a^2-\frac{2GMr}{c^2}
=\left(1-r_-\right)\left(1-{r_{+}}\right)=0.
\quad
\label
{Delta=0}
\end{equation}                                          
If $a=GM/c^2$, then there is a single event horizon.
If $a>GM/c^2$, then the  event horizon is nonexistent, and 
the curvature singularity is bare.

Thus, the analysis becomes more involved when the black hole rotation
is incorporated. 
Things get worse in particular 
because there is no known explicit
solution with rotating matter whose final state is described by a stationary
geometry with the Kerr metric.

Let us now pass over charged black holes described by the Reissner--Nordstrom
and Kerr--Newman solutions. 
Despite further technical subtleties, the major conclusion that 
such objects are suitable to studying a classical-quantum phase 
transition
still stands.

We close this section by noting that a black hole horizon shows a clear
demarcation boundary between spacetime regions characterized by opposite signatures.
It seems likely that {\bf this geometrical layout, if it exists, 
is the only 
opportunity for interfacing 
the classical and the quantum}.

\section{The Foldy--Wouthuysen picture}
A convenient framework for analyzing the classical and quantum regimes of
evolution from  a unified perspective is provided by the Dirac equation.
Foldy and Wouthuysen \cite{Foldy--Wouthuysen} found 
a unitary transformation which diagonalizes the free Dirac 
Hamiltonian $H_0$ with respect to positive and negative energies,
\begin{equation}
U_{\rm FW}^{-1}H_0U_{\rm FW}
=
                   \pmatrix{\sqrt{-\nabla^2+m^2} &0\cr
                            {}\,\,\,\,\,\,0            &-\sqrt{-\nabla^2+m^2}       \cr}
=
\beta\,|H_0|.
\label
{H-Dirac-to-H-FW}
\end{equation}                                          
From here on we will use the natural system of units in which $c=1$ and $\hbar=1$.
The standard representation of Dirac matrices is assumed,
\begin{equation}
\alpha^i
=
                   \pmatrix{0        &\sigma^i\cr
                            \sigma^i &0       \cr},
\quad
\beta
=
                   \pmatrix{I        &{}\,\,\,0\cr
                            0        &-I       \cr},
\label
{alpha-beta-matrices}
\end{equation}                                          
where $I$ is the $2\times 2$  unit matrix, and $\sigma^i$ are Pauli matrices. 

Following the conventional interpretation, we consider 
positive energy states as those attributed to a Dirac particle,
while states of negative energy are those attributed to its
antiparticle.
Thus, the free Dirac equation
\begin{equation}
i\,\frac{\partial}{\partial t}\,\psi=\left[i\left({\vec\alpha}\cdot\nabla\right)
+\beta m \right]\psi
\label
{Dirac-eq}
\end{equation}
is unitarily equivalent to a pair of 
two-component equations
\begin{equation}
i\,\frac{\partial}{\partial t}\,\chi=\beta\sqrt{-\nabla^2
+m^2}\,\chi.
\label
{FW-eq}
\end{equation}

An important point is that a separation of positive- and negative-energy states
is still possible in the presence of an external
time-independent magnetic field of arbitrary strength ${\bf B}$.
Case \cite{Case} obtained a closed form for the Foldy--Wouthuysen transformation 
of the Dirac equation into  
\begin{equation}
i\,\frac{\partial}{\partial t}\,\chi=\beta\sqrt{(i\nabla-e{\bf A})^2
-e\,({\vec\sigma}\cdot{\bf B})+m^2}\,\chi,
\label
{FW-Case-eq}
\end{equation}
where ${\bf A}$ is the vector potential of this field,  
${\bf B}={\rm curl}\,{\bf A}$.
The energy gap of the Dirac sea is not penetrated in
the constant
magnetic field because this field leaves the energy of the Dirac particle unchanged.  

However, the separation 
is not possible with
time-dependent electromagnetic fields and scalar potentials;
the positive- and negative-energy solutions may mix when the interaction is
sufficiently strong
(a general review of these and related problems can be found in 
\cite{deVries}--\cite{Neznamov}). 
The energy gap of the Dirac sea is no longer
insuperable.
Creations and annihilations are made possible.

We interpret these facts by saying that 
a Dirac particle manifests itself as a classical entity in the case that the
Hamiltonian can be 
diagonalized  with respect to positive and negative energies.

Is it possible to transform the Dirac 
equation in a curved spacetime 
to the Foldy--Wouthuysen form?
Obukhov \cite{Obukhov} showed that 
an exact
Foldy--Wouthuysen transformation $U_{\rm FW}$ can be written for stationary 
metrics 
\begin{equation}
ds^2=V^2({\bf r})\,dt^2-W^2({\bf r})\,d{\bf r}^2,
\label
{stationary-metrics}
\end{equation}
where $V$ and $W$ are arbitrary functions of spatial coordinates
${\bf r}=(x,y,z)$.
Schwarzschild geometry is a particular case of (\ref{stationary-metrics}).
When employing isotropic coordinates, the metric 
(\ref{Schwarzschild-metric}) takes the form
(\ref{stationary-metrics}) with
\begin{equation}
V=\left(1-\frac{r_{\rm S}}{4r}\right)\left(1+\frac{r_{\rm S}}{4r}\right)^{-1},
\quad
W=\left(1+\frac{r_{\rm S}}{4r}\right)^2.
\label
{V-W-Schwarzschild}
\end{equation}

To be more specific, we recall that Dirac particles in curved backgrounds
are governed by the covariant Dirac equation
\begin{equation}
\left(i\gamma^{\hat\alpha} D_{\hat\alpha} -m\right)\psi=0.
\label
{covariant-Dirac-eq}
\end{equation}
Here, $D_\alpha$ is the spinor covariant derivative
\begin{equation}
D_{\hat\alpha}=e^\mu_{~{\hat\alpha}}\left(D_\mu+\frac{1}{4}\,\Gamma_\mu\right),
\label
{covariant-derivative}
\end{equation}
with the  gravitational gauge potential $\Gamma_\mu=\frac14\,
[\gamma^{\hat\alpha},\gamma^{\hat\beta}]\,\gamma_{{\hat\alpha}{\hat\beta}
{\hat\gamma}}\,e^{\hat\gamma}_{~\mu}$
and the Ricci rotation coefficients $\gamma_{{\hat\alpha}{\hat\beta}{\hat\gamma}}=
e_{\hat\alpha}^{~\mu}\,e_{{\hat\beta}\mu;\nu}\,e_{\hat\gamma}^{~\nu}$.
If (\ref{covariant-Dirac-eq}) can be brought into 
the  form
\begin{equation}
i\,\frac{\partial\psi}{\partial t}= H\psi
\label
{Dirac-eq-in-Schroedinger-form}
\end{equation}
with
\begin{equation}
H=\beta m-\frac{i}{2}\left[({\vec\alpha}\cdot\nabla)F+
F({\vec\alpha}\cdot\nabla)\right],
\quad
F=\frac{V}{W},
\label
{Hamiltonian-Obukhov-form}
\end{equation}
then there is a unitary transformation $H_{\rm FW}=
U_{\rm FW}HU_{\rm FW}^\dagger$ such that
\begin{equation}
H_{\rm FW}=
\frac{1}{2}\left(\sqrt{{\bar H}^2}
+\beta\sqrt{{\bar H}^2}\beta\right)
+\frac{1}{2}\left(\sqrt{{\bar H}^2}-\beta\sqrt{{\bar H}^2}\beta\right) 
\gamma_5\beta,
\label
{FW-Hamiltonian-Obukhov}
\end{equation}
\begin{equation}
{\bar H}^2
=m^2V^2
-F\,\nabla^2\,F
+\frac{1}{2}\,F\,(\nabla{\vec f})-\frac{1}{4}\,{\vec f}^2
-i\,F\,{\vec\Sigma}\cdot\left[{\vec f}\times\nabla
-\gamma_5\beta m\,{\vec\phi}\,\right].
\label
{FW-Hamiltonian-sqr-Obukhov}
\end{equation}
Here, ${\vec\Sigma}$ is the spin matrix ${\vec\Sigma}=\frac12\,i\,{\vec\gamma}
\times{\vec\gamma}$, and ${\vec f}= \nabla F$, ${\vec\phi}= \nabla V$.

A remarkable fact is that the feasibility
of an exact Foldy--Wouthuysen transformation is another way of
stating that the system enjoys the property of supersymmetry \cite{Zentella}.
The origin of the Foldy--Wouthuysen picture for a Dirac 
particle in an external electromagnetic field is related to the existence of
a supercharge. 
If 
the Dirac sea is stable,
then the positive- and negative-energy solutions are supersymmetric 
partners of each other.
On the other hand, when the supersymmetry is broken, it is impossible to obtain
an exact block-diagonalized Hamiltonian for this system. 

These findings can be extended to curved spacetimes \cite{Heidenreich}.
A supercharge can be constructed for a
relatively wide class of stationary metrics, including that defined in  
(\ref{stationary-metrics}).
The  Foldy--Wouthuysen transformed Hamiltonian $H_{\rm FW}$ is proportional
to the square root of the super-Hamiltonian 
\begin{equation}
H_{\rm FW}=\beta\sqrt{Q^2}.
\label
{H-FW=sqrt-Q-2}
\end{equation}
For the metric (\ref{stationary-metrics}), 
\begin{equation}
Q=
\frac{1}{2}\left\{{\vec\alpha}\cdot p,F\right\}+JV.
\label
{Q-Obukhov}
\end{equation}
Here, $J$ is the involution operator $J=i\gamma_5\beta$
(a Hermitian and unitary operator, $J^\dagger=J$, $JJ^\dagger=1$, which
anticommutes with both the Hamiltonian and the $\beta$ matrix,
$JH+HJ=0$, $J\beta+\beta J=0$).

We now come to the following problem.
Consider a self-gravitating Dirac field $\psi$ which 
arranges itself into a spherically symmetric 
{collapsing wave packet}.
Let the total mass of the wave packet be equal 
to $M$, the parameter which enters in the definition of the Schwarzschild 
radius $r_{\rm S}$.
Before a black hole state settles down, the $\psi$ is assumed to
model an astrophysical collapsing object in the Schwarzschild spacetime.
Our interest here is with the `last stage' of evolution 
of the wave packet just before its shrinking down to below the horizon.
At this stage, the $\psi$ is governed by the diagonalized Foldy--Wouthuysen 
Hamiltonian (\ref{FW-Hamiltonian-Obukhov}).
It would be desirable to have an exact solution to this problem.
Perhaps this solution will exhibit a singular point after which this 
classical regime of evolution is
no longer valid.
Note that this singularity is just the point in which the supersymmetry must be violated.

Of course a similar discussion of rotating spacetimes, say, based on the Kerr
geometry, would be highly desirable.
This would provide insight into the mechanism of forming a black
hole from a collapsing fluid $\psi$ endowed with spin degrees of freedom.

\section{The classical path integral}
Not only quantum mechanics, but also classical mechanics can be treated 
through a path 
integral formulation proposed by Gozzi \cite{gozzi}, and developed in a
long series of papers by Gozzi and his collaborators (see \cite{grt, persik} and references therein).
We now outline the basic elements of this treatment.

The probaility amplitude 
of finding a classical system at a phase space point 
$\phi^a_f=(q^a_f,p^a_f)$ at time $t_f=T$ if it was at 
$\phi^a_i=(q^a_i,p^a_i)$ at
time $t_i=0$
is given by 
\begin{equation}
K(\phi^a_f, T|\phi^a_i,0)=\int \left[{{\cal D}}\phi\right]\delta\left[\phi^a-
\phi^a_{\rm cl}\left(T;\phi_i,0\right)\right].
\label
{gozzi---PI}
\end{equation}                                           
Here, $\phi^a_{\rm cl}$ is the solution to the classical equation of motion
${\dot\phi}^a=\omega^{ab}\partial_bH$, with $\omega^{ab}$ being a symplectic 
matrix, and $H$ the Hamiltonian of this system.
The symbol $[{{\cal D}}\phi]$ indicates that the integration is over
the all phase space paths with fixed end points ${\phi}_i$ and ${\phi}_f$.
 
Since
\begin{equation}
\delta(\phi-\phi_{\rm cl})
=
\delta\!\left({\dot\phi}^a-\omega^{ab}\partial_bH\right)
\det\left(\delta^a_{~b}{\partial}_t-\omega^{ac}\partial_c{\partial}_bH
\right),
\label
{delta=delta-det}
\end{equation}        
one may further take the Fourier transform of the Dirac delta and exponentiate
the determinant using an even Grassmannian variable $\lambda_a$ and odd 
variables $c^a$ and ${\bar c}_a$
to yield
\begin{equation}
K(\phi_f,T|\phi_i,0)
=
\int [{{\cal D}}\phi]\,
{\cal D}\lambda\,{\cal D}c\,{\cal D}{\bar c}\,
\exp\!\left(i\int_0^T dt\, {\widetilde{\cal L}}\right).
\label
{gozzi-PI}
\end{equation}                                           
where
\begin{equation}
{\widetilde{\cal L}}
=
\lambda_a{\dot\phi}^a
+i{\bar c}_a{\dot c}^a
-\lambda_a \omega^{ab}\partial_bH
-i{\bar c}_a\omega^{ad}\partial_d\partial_bH c^b.
\label
{gozzi-Lagrangian}
\end{equation}                                           

If we define two anticommuting partners of $t$, ${\bar\theta}$ and $\theta$, and
assemble the variables $\phi,\lambda,{\bar c},c$ into a single 
combination of supersymmetric phase space coordinates
\begin{equation}
Q(t,\theta,{\bar\theta})
=q(t)
+{\theta}{c}^q
+{\bar\theta}{\bar c}_p
+i{\bar\theta}\theta\lambda_p,
\label
{super-Q}
\end{equation}                                           
\begin{equation}
P(t,\theta,{\bar\theta})
=p(t)
+{\theta}{c}^p
-{\bar\theta}{\bar c}_q
-i{\bar\theta}\theta\lambda_q,
\label
{super-P}
\end{equation}                                           
then it is possible to rewrite (\ref{gozzi-PI}) in a very compact and 
elegant form \cite{persik}:
\begin{equation}
K(Q_f,T|Q_i,0)=\int \left[{{\cal D}}Q\right]{\cal D}P\,
\exp\!\left[i\int_0^T idt\,d\theta\,d{\bar\theta}\,\,L(Q,P)\right],
\label
{gozzi-PI-final}
\end{equation}                                           
where $L$ is the usual Lagrangian of this system $L(q,p)=p{\dot q}-H(q,p)$.
Equation (\ref{gozzi-PI-final}) bears the formal similarity to the 
quantum path integral
\begin{equation}
K(q_f,T|q_i,0)
=\int \left[{\cal D} q\right] {\cal D}p\,\exp\left[\frac{i}{\hbar}\int_0^T
dt\,L(q,p)\right].
\label
{path-int-QM}
\end{equation}
In fact, (\ref{gozzi-PI-final}) derives from (\ref{path-int-QM}) by
replacing the phase space coordinates $q,p$ with the super phase space coordinates 
$Q,P$ and extending the integration over $t$ to an integration over the
supertime $(t,\theta,{\bar\theta})$, multiplied by $\hbar$.

Turning back to our main subject, 
consider a gravitating perfect fluid in a Schwarzschild
background, which arranges itself into a 
collapsing ball.
The canonical theory of classical perfect fluids, 
in Eulerian and Lagrangian formulations, is well studied (a review
can be found in \cite{Jackiw}).
It is then desirable to extend this theory to curved spacetimes, construct
its supersymmetric version by substituting the phase space for the super 
phase space, as shown in 
(\ref{super-Q})--(\ref{super-P}), and write 
the classical path integral (\ref{gozzi-PI-final}).
If we would succeed in working out this integral, then we would read from 
the resulting expression a
self-denial of the classical physics at some point.
Notice that the supersymmetry structure of equation (\ref{gozzi-PI-final}) 
is automatically broken at
this point.

\section{Conclusion}
Let us summarize the results of our discussion.

The black hole formation bears some resemblance to a {phase transition}.
The approach to the black-hole state occurs fairly {quickly}.
The {energy content} of the system in an initial state
differs significantly from that after the hole has formed.
{This} transition entails a profound {symmetry 
rearrangement}:
by the no hair theorems, all initial symmetries  
and their associated conserved quantities, except for $M$, $J$, and 
$Q$, disappear in
the ultimate black-hole state. 

The central idea of this paper is that
{forming the black hole horizon implies a
transition between the classical and quantum regimes of evolution}. 

When the essentials of classical 
theory is compared with those of quantum theory, 
it transpires that {spontaneous creations and annihilations of 
particle-antiparticle pairs are impossible in the classical world, but 
possible in the quantum
world}.
Furthermore,
it is sufficient to 
{change the overall sign of the
spacetime signature in the classical description of field propagation for
it to be treated as the quantum description of field propagation}.
This leads us to consider the black hole horizon as a sharply defined
boundary which demarcates the classical and the quantum.

The feasibility
of the classical regime of evolution 
is another way of
stating that the system enjoys the property of {supersymmetry}.
To describe a self-gravitating object
at 
the last stage of its classical evolution, 
just before its shrinking down to below the horizon, 
we proposed to invoke the
Foldy--Wouthuysen 
representation of the Dirac equation in curved spacetimes,
and/or the Gozzi classical path integral technique.
In both descriptions, maintaining the dynamics in the classical 
regime is controlled 
by supersymmetry.
If we would succeed in 
integrating the
Foldy--Wouthuysen dynamics for a collapsing {wave packet}  $\psi$
(say, in the Schwarzschild background),
or, alternatively,
in calculating the Gozzi path integral for a 
gravitationally collapsing perfect fluid, we would establish a point 
where a self-destruction of this classical machinery occurs.
The supersymmetry structures undergoes a  breakdown at
this point.

\section*{Acknowledgments}
I thank the organizers of the Tomsk QFT{\&}G  conference for the pleasant and
stimulating working atmosphere.
I am grateful to Dieter Brill for his illuminating remarks.

\end{document}